\documentclass[prl,aps,twocolumn]{revtex4}

\usepackage{graphicx,amssymb,amsmath,color,psfrag}
\usepackage{amsthm}
\usepackage{amsfonts}
\usepackage{algorithmic}
\usepackage{enumerate}
\usepackage{latexsym}

\begin{document}

\title{Electronic band gaps and transport in aperiodic graphene
superlattices of Thue-Morse sequence}
\author{Tianxing Ma$^{1,2,}$\footnote{txma@bnu.edu.cn}, Chun Liang$^{2}$, Li-Gang Wang$^{3,2,}$\footnote{sxwlg@yahoo.com.cn} and
Hai-Qing Lin$^{2}$}
\affiliation{$^{1}$Department of Physics, Beijing Normal University, Beijing 100875, China%
\\
$^{2}$Beijing Computational Science Research Center, Beijing 100084, China\\
$^{3}$Department of Physics, Zhejiang University, Hangzhou 310027, China\\
}

\begin{abstract}
We have studied the electronic properties in aperiodic graphene
superlattices of Thue-Morse sequence. Although the structure is aperiodic,
an unusual Dirac point (DP) does exist and its location is exactly at the
energy corresponding to the zero-averaged wave number (zero-$\overline{k})$.
Furthermore, the zero-$\overline{k}$ gap associated with the DP is robust
against the lattice constants and the incident angles, and multi-DPs may
appear under the suitable conditions. A resultant controllability of
electronic transport in Thue-Morse sequence is predicted, which may facilitate
the development of many graphene-based electronics.
\end{abstract}

\maketitle


Graphene has attracted enormous attention of experimentalists and theorists
\cite%
{Novoselov2004,Zhang2005,Kuzmenko2008,Wang2008,XChen2009,reviews,TxMa2010,Peres2010}
since its discovery.
The interest is driven by its potential technological applications and
unconventional low-energy behavior, since graphene has a unique band
structure with the conductance and valance bands touching at Dirac point
(DP). Recently, scientists anticipate that graphene-based optoelectronics
may supplement silicon-based technology, which is nearing its limits \cite%
{SAWolf2001}. For superlattices are vastly successful to control the
electronic transport \cite{Tsu2005}, to facilitate the application of
graphene-based devices, the graphene superlattices (GSLs) with electrostatic
potential or magnetic barrier have also 
received broad focuses of theoretical and experimental investigations\cite%
{JCM2008,SMarchini2007,ALVa2008,CXBa2007,MBarb2010,LBrey2009,CHP2008,LGWang2010,XXGuo2011,XiChen2011}%
. In such GSLs, the new DP appears in the band structures \cite%
{MBarb2010,LBrey2009} and it's exactly located at the energy corresponding
to zero-averaged wave number (zero-$\overline{k}$) \cite{LGWang2010}.
Contrary to Bragg gaps, the zero-$\overline{k}$ gap associated with the new
DP is insensitive to both the lattice constant and the structural disorder,
resulting in better controllability of electronic transport in GSLs. Most
recently, Zhao and Chen predicted a controllable electron transport in a
Fiboncacci quasi-periodic GSL \cite{XiChen2011}.

In this letter, we investigate the electronic band gaps and transport in the
graphene-based Thue-Morse (TM) sequence. As a typical aperiodic system, the
TM lattice has been widely studied\cite%
{MQu1987,ZCheng1988,Axel1989,Luck1989,Jiang2005,Noh2011}, which is known to
have a singular continuous Fourier transform\cite{ZCheng1988} and a
Cantor-like phonon spectrum\cite{Axel1989}, and it is more \textquotedblleft
disordered\textquotedblright\ than the Fibonacci sequence \cite%
{ZCheng1988,MQu1987}. The TM lattice has a deterministic geometry structure,
and its electronic properties are also of great interesting\cite{Luck1989}.
In the studied graphene-based TM sequence, we find that zero-$\overline{k}$
gap and DP do exist, which results in robust electronic transport
properties. 
Moreover, the splitting behavior of the passing bands in
the graphene-based TM sequence is very different from that in the
graphene-based Fibonacci sequence, which leads to the different electronic
transport property.

\begin{figure}[tbp]
\includegraphics[scale=0.80]{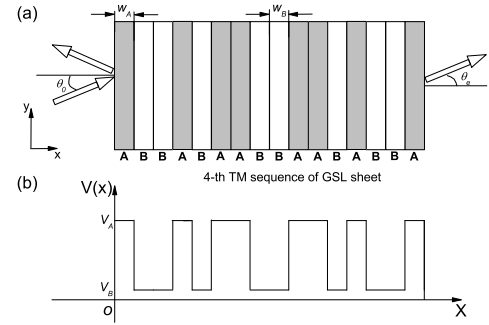}
\caption{(Color online) (a) Example of a 4-th order GSL TM sequence. (b) The schematic profiles of the potentials $V_{A}$ and $V_{B}$ corresponding to (a). }
\label{Fig:TM}
\end{figure}

For an $n$-th TM sequence, $S(n)$, it contains elements $A$ and $B$, and
follows the inflation rule: $A\rightarrow AB,\ B\rightarrow BA$ from
generation to generation with $S(1)=AB$ (for example, see Ref. \cite%
{NhLiu1997}). Let $S_{1/2}^{L}(n)$ and $S_{1/2}^{R}(n)$ denote the left and
right half parts of an $n$-th order TM sequence, respectively, then
iteration relation for any $(n+1)$-th TM sequence is written as
\begin{equation}
S(n+1)=S(n)\ S_{1/2}^{R}(n)\ S_{1/2}^{L}(n).  \label{TMrule}
\end{equation}%
Naturally, $S(2)=ABBA$, $S(3)=ABBABAAB$, and so on. In our cases, $A$ ($B$)
denotes barrier $V_{A}$ ($V_{B}$) with its width $w_{A}$ ($w_{B}$). 
Fig. 1 shows the schematic illustration of the 4th order GSL TM
sequence and the corresponding distributions of barriers and wells.
\textit{From Eq. (\ref{TMrule}) and Fig. 1}, it is readily to know that the
number of $A$, $N_{A}$, equals to that of $B$, $N_{B}$, in any $S(n)$, i.e.,
$N_{A}=N_{B}=2^{n-1}$.

\begin{figure}[tbp]
\includegraphics[scale=0.6]{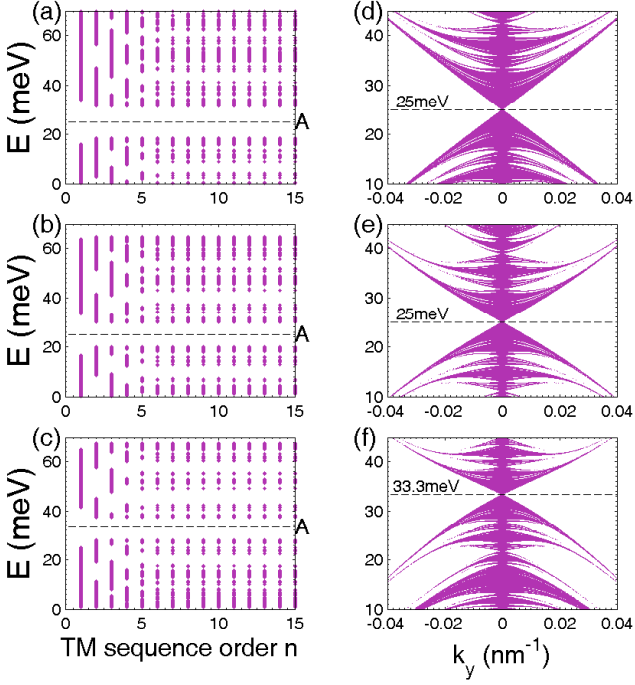}
\caption{(Color online) Trace-maps for GSL TM sequences of variable order $n$
under (a) $w_{A,B}=20$ nm, (b) $w_{A,B}=25$ nm and (c) $w_{A}=2w_{B}=40$ nm
with $V_{A}=50$ meV, $V_{B}=0$, and $k_{y}=0.015$ nm$^{-1}$. (d), (e) and
(f) are band structures for $n=8$ corresponding to (a), (b) and (c),
respectively. The horizontal dashed black line denotes the new Dirac point's
location, and the white areas are band gaps and the solid areas are passing
bands. }
\label{Fig:fig2}
\end{figure}

Meanwhile the charge carriers near the \emph{\textbf{K}} point in graphene
(near the Fermi level) are described by the Hamiltonian: $\widehat{H}%
=\upsilon _{F}\widehat{\mathbf{\sigma }}\cdot \widehat{\mathbf{p}}+V(x)%
\widehat{I}$, where $V(x)$ is a barrier or well, $\upsilon _{F}\approx
10^{6} $m/s is the Fermi velocity, $\widehat{\mathbf{\sigma }}=\left(
\widehat{\sigma }_{x},\widehat{\sigma }_{y}\right) $ are Pauli matrices, and
$\widehat{I}$ is a $2\times 2$ unit matrix. The solution of $\widehat{H}$,
acting on the electronic pseudospin wavefunctions, leads to a transfer
matrix \cite{LGWang2010}
\begin{equation}
M_{j}(\Delta x,E,k_{y})=\left(
\begin{array}{cc}
\frac{\cos (q_{j}\Delta x-\theta _{j})}{\cos \theta _{j}} & i\frac{\sin
(q_{j}\Delta x)}{\cos \theta _{j}} \\
i\frac{\sin (q_{j}\Delta x)}{\cos \theta _{j}} & \frac{\cos (q_{j}\Delta
x+\theta _{j})}{\cos \theta _{j}}%
\end{array}%
\right) \ ,  \label{trans}
\end{equation}%
which connects the wave functions at $x$ and $x+\triangle x$ inside the $j$%
th potential with $\theta _{j}=\text{arcsin}(k_{y}/k_{j})$, here $k_{y}$ is
the $y$ component of $k_{j}$, and $q_{j}=\text{sign}(k_{j})\sqrt{%
k_{j}^{2}-k_{y}^{2}}$ for $k_{y}^{2}<k_{j}^{2}$, otherwise $q_{j}=i\sqrt{%
k_{y}^{2}-k_{j}^{2}}$ for $k_{y}^{2}>k_{j}^{2}$. The electronic transmission
coefficient $t(E,k_{y})$ in such devices can be obtained by \cite{LGWang2010}
\begin{equation}
t(E,k_{y})=\frac{2\cos \theta _{0}}{(x_{22}e^{-i\theta
_{0}}+x_{11}e^{i\theta _{e}})-x_{12}e^{i(\theta _{e}-\theta _{0})}-x_{21}},
\end{equation}%
where $\theta _{0}$($\theta _{e}$) is the incident (exit) angle, and $%
x_{ij}(i,j=1,2)$ is the element of $\mathbf{X}[S(n)]=\prod%
\limits_{j=1}^{N}M_{j}(w_{j},E,k_{y})$, which is the entire transfer matrix
of a TM sequence. If we let $x_{n}=\mathbf{Tr\{X[}S(n)]\}$, it is easy to
derive the iteration relation for the trace map of the $n$-th GSL TM
sequence as follows \cite{MKol1991,Axel1986,NhLiu1997}:
\begin{equation}
x_{n}=x_{n-2}^{2}\left( x_{n-1}-2\right) +2.\   \label{Xn}
\end{equation}%
Treating an $n$-th TM sequence as a unit cell, from Bloch's theorem, we have
$\cos \left( \beta _{x}\Lambda _{n}\right) =x_{n}/2$, where $\Lambda
_{n}=N_{A}w_{A}+N_{B}w_{B}$. From Eq.(\ref{Xn}), we can calculate the change
of the trace map as a function of the order $n$.

In Fig. 2 (a) and (b), we plot the trace maps for two kinds of
graphene-based TM sequences with the change of $n$, at the incident angle $%
\theta _{0}=10^{\circ }$. We take $w_{A,B}=20$nm in Fig. 2(a), and $%
w_{A,B}=25$nm in Fig. 2(b). We find that the passing bands are split into
more and more sub-bands as $n$ increases; when $n\geq 6,$ the band
structures almost become the discontinuous bands. However, we note that the
center positions of the Gap A in Figs. 2(a) and 2(b) are the same, and the
positions of other gaps are shifted with the change of the lattice constant.
Corresponding to Fig. 2(a) and 2(b), the electronic band structures for a TM
sequence of $n=8$ have been shown in Fig. 2(d) and 2(e), respectively. It is
seen that a new DP appears inside the Gap A and this new DP actually locates
at
\begin{equation}
\overline{k}=\sum\limits_{j=1}^{N}k_{j}w_{j}/\sum\limits_{j=1}^{N}w_{j}=0.
\label{SK1}
\end{equation}%
This condition is valid for periodic and aperiodic graphene superlattices
\cite{LBrey2009,LGWang2010,XiChen2011}. Since $N_{A}=N_{B}$ in our cases, it
is easily to obtain the energy for $\overline{k}=0$ as follows:%
\begin{equation}
E=\frac{V_{A}+V_{B}\cdot w_{B}/w_{A}}{1+w_{B}/w_{A}}.  \label{Energy}
\end{equation}%
This condition is different from the case of the Fibonacci sequence,
which depends on the ration of numbers of layer A and B \cite{XiChen2011}.
Actually, our formula (\ref{SK1}) is valid for the condition of the new DP
in the graphene-based Fibonacci sequence. Here we would like to emphasize that
for the TM sequence, the location of the DP is independent of order $n$;
while for the Fibonacci sequence it changes for different order $n$.
As an example for the TM sequence, in cases of Fig. 2(a,b,d,e), it is given by $E=25$meV since $%
w_{A}=w_{B}$.
\begin{figure}[tbp]
\includegraphics[scale=0.4]{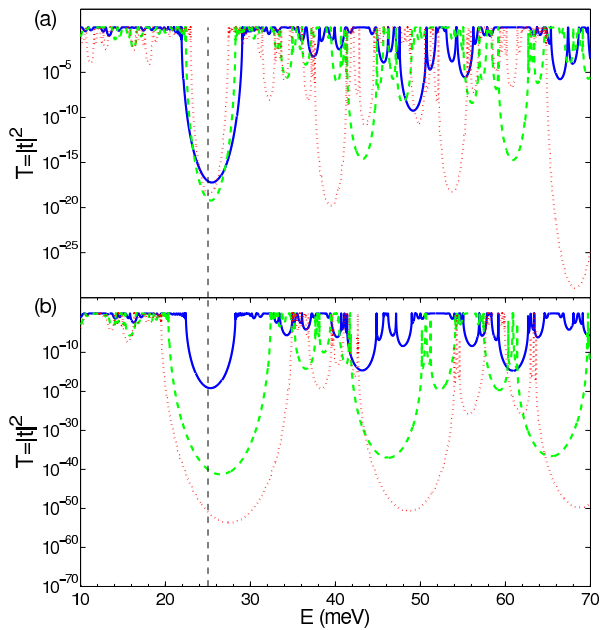}
\caption{(Color online) (a) Dependence of transmission spectrum (at $\protect%
\theta _{0}=10^{\circ }$) on different widths of barriers and wells, here
solid blue line for $w_{A}=w_{B}=15$nm, dashed green line for $%
w_{A}=w_{B}=20 $nm, and dotted red line for $w_{A}=w_{B}=25$nm; (b)
dependence of transmission spectrum on different incident angles: $\protect%
\theta _{0}=10^{\circ }$ (solid blue line), $20^{\circ }$ (dashed green),
and $25^{\circ }$ (dotted red) with fixed lattice constants $w_{A}=w_{B}=20$%
nm. Other parameters are $V_{A}=50$meV and $V_{B}=0$ for an 8-th GSL TM
sequence with total 256 layers. }
\label{Fig:fig3}
\end{figure}

Since the center of the Gap A in Fig. 2 is located at zero-$\overline{k}$,
which is denoted by a dashed black line in Fig. 2(a) and 2(b), we may call
it as the zero-$\overline{k}$ gap. According to Eq.(\ref{Energy}), the
position of the zero-$\overline{k}$ gap for the TM sequences is not shifted
with the lattice constant itself, but is shifted with the change of the
ratio of $w_{A}/w_{B}$. As shown in Fig.2(c) and (f), the position of the
zero-$\overline{k}$ gap and the DP move to $E\approx 33.33$meV when $%
w_{A}/w_{B}=2$.

Fig.3 (a) shows the effect of the lattice constants on the electronic
transmission spectrum. Based on Fig. 3(a), it is obvious that the zero-$%
\overline{k}$ gap is insensitive to the lattice constants themselves.
However, the positions of other gaps and passing bands with higher energy
are highly dependent on the lattice parameters. From Fig. 3(b), one can also
find that the position of zero-$\overline{k}$ gap is weakly dependent on the
incident angle $\theta _{0}$, while other Bragg gaps change sensitively with
$\theta _{0}$.

\begin{figure}[tbp]
\includegraphics[scale=0.425]{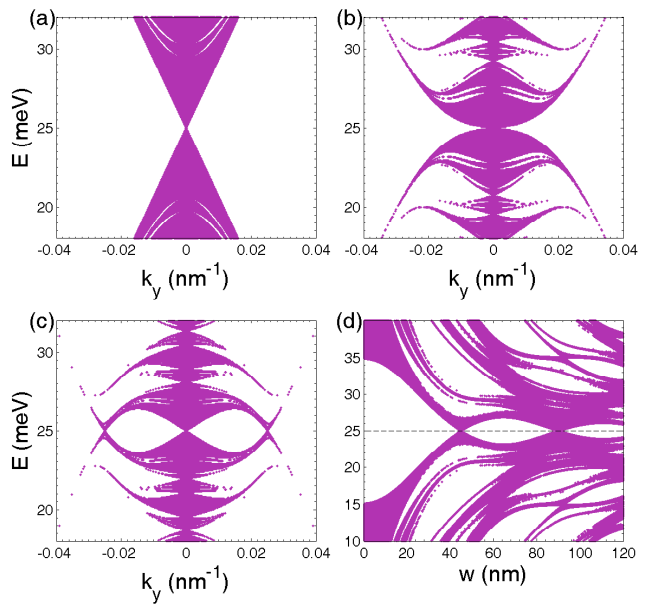}
\caption{(Color online) Electronic band structures for (a) $w_{A}=w_{B}=20$%
nm, (b) $w_{A}=w_{B}=40$nm and (c) $w_{A}=w_{B}=55$nm; (d) band gaps
depending on the lattice constant $w$ with fixed $k_{y}=0.015$ nm$^{-1}$,
here $w_{A}=w_{B}=w$. Other parameters are the same as those in Fig.2. }
\label{Fig:fig4}
\end{figure}

Furthermore, one has known that the multi-Dirac-points could appear in the
GSLs with periodic potential structures \cite{LBrey2009,LGWang2010}. Here we
point out that the extra Dirac points, located at $k_{y}\neq 0$, could also
emerge in the GSL TM sequence as the lattice constant increases. See Figs.
4(a) and (b), the slope of the band edges near the DP gradually turns
smaller as $w_{A}$ ($w_{B}$) increases from 20 nm to 40 nm. When $w_{A}$ ($%
w_{B}$) is larger than 40 nm, as that shown in Fig. 4 (c), the additional
Dirac points appear at the same energy. From Figs. 4(a) to 4(c), we may also
find the appearance of additional Dirac points associating with the
variation of the zero-$\overline{k}$ gap. In Fig. 4 (d), we show the energy
band gaps as a function of the lattice constants in the case of $w_{A}=w_{B}$%
. In the process of the appearance of the additional Dirac points, the zero-$%
\overline{k}$ gap opens and closes oscillationally while the other gaps are
shifted greatly.

\begin{figure}[tbp]
\includegraphics[scale=0.45]{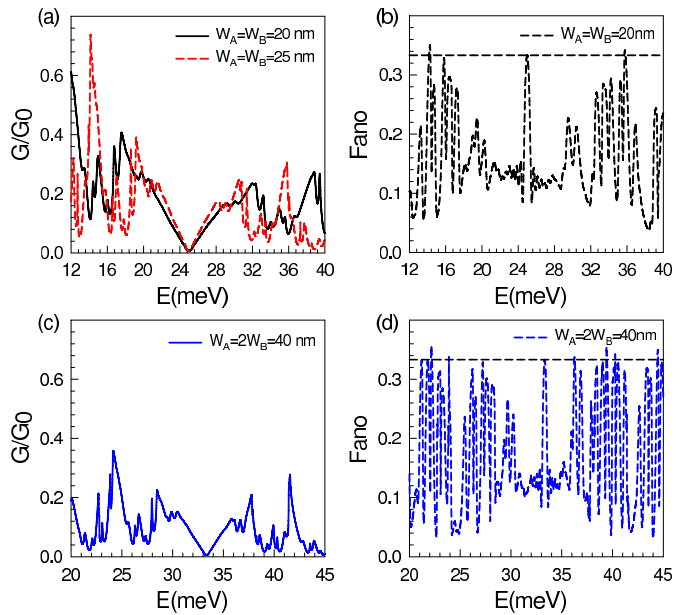}
\caption{(Color online) Conductance [(a) and (c)] and Fano factor [(b) and
(d)] vs Fermi energy in an 8-th GSL TM sequence. Other parameters are the
same as those in Fig.2.}
\label{Fig:fig5}
\end{figure}

Finally, we calculate the total conductance $G$ \cite{SDatta1995} and the
Fano factor $F$ \cite{JTwor2006} in GSL TM sequences, which are given by $G$=%
$G_{0}\int\nolimits_{0}^{\pi /2}T\cos \theta _{0}d\theta _{0}$ and $F$=$%
\int\nolimits_{-\pi /2}^{\pi /2}T\left( 1-T\right) \cos \theta _{0}d\theta
_{0}/\int\nolimits_{-\pi /2}^{\pi /2}T\cos \theta _{0}d\theta _{0},$ where $T
$=$|t|^{2}$ and $G_{0}$=$2e^{2}m\upsilon _{F}L_{y}/\hbar ^{2},$ with $L_{y}$
denoting the width of the graphene stripe in the $y$ direction. In Fig. 5, we
present the total conductance and the Fano factor as a function of Fermi
energy with different lattice constants. What we should notice is that the
angular-averaged conductance $G$'s curve reaches its minimum at the DP and
forms a linear cone around the DP, and $F$ at the DP's location reaches the
value of $1/3$ approximately \cite{XXGuo2011,JTwor2006}. From Fig. 5, we
also find that the conductance and the Fano factor shift with the ratio of $%
w_{A}/w_{B}$ since the DP's location relies on this ratio. Therefore, this
indicates that the conductance of GSL TM sequence could be modulated by the
ratio of the lattice constants. 
 In addition, we should point out
that, if one compares the passing bands of Fig. 2(a) in our case with those
of Fig. 2(a) in Ref. \cite{XiChen2011}, one can easily find that the
splitting behavior of bands in the TM sequence is very different from that
in the Fibonacci case. Correspondingly, the electronic transport properties
(the values of G and F) are different from those in the Fibonacci case.
These distinct differences between the TM and Fibonacci sequences will be
presented in our future work.

In summary, we have studied the electronic transport properties in the
graphene-based Thue-Morse aperiodic sequence. It is shown that the extra-DPs
can appear under some suitable conditions. 
The zero-$\overline{k}$ gap associated with the DP is robust against the
lattice constants and the incident angles, and the resultant controllable
electron transport may facilitate the development of graphene-based
electronics.

\begin{acknowledgments}
This work is supported by NSFCs (Grant. No. 11104014 and No. 61078021),
Research Fund for the Doctoral Program of Higher Education of China
20110003120007 and the National Basic Research Program of China (Grant No.
2012CB921602).
\end{acknowledgments}

\end{document}